\documentclass{INTERSPEECH2023}

\interspeechcameraready

\title{Learning to Compute the Articulatory Representations of Speech with the MIRRORNET}
\name{Yashish M. Siriwardena, Carol Espy-Wilson, Shihab Shamma}

\address{Institute for Systems Research, University of Maryland College Park, USA}
\email{yashish@umd.edu, espy@umd.edu, sas@umd.edu}

\begin{document}

\maketitle
 
\begin{abstract}
Most organisms including humans function by coordinating and integrating sensory signals with motor actions to survive and accomplish desired tasks. Learning these complex sensorimotor mappings proceeds simultaneously and often in an unsupervised or semi-supervised fashion. An autoencoder architecture (MirrorNet) inspired by this sensorimotor learning paradigm is explored in this work to control an articulatory synthesizer, with minimal exposure to ground-truth articulatory data. The articulatory synthesizer takes as input a set of six vocal Tract Variables (TVs) and source features (voicing indicators and pitch) and is able to synthesize continuous speech for unseen speakers. We show that the MirrorNet, once initialized (with $\sim$30 mins of articulatory data) and further trained in unsupervised fashion (`learning phase'), can learn meaningful articulatory representations with comparable accuracy to articulatory speech-inversion systems trained in a completely supervised fashion. 


\end{abstract}

\vspace*{-6pt}
\section{Introduction}
\vspace*{-2pt}

Advances in Deep Neural Networks (DNNs) have significantly contributed to learning complex mappings and interactions between perception and action domains. These models are extensively trained on large databases that map the input sensory data to their corresponding actions \cite{imitation_RL_for_robots}. Humans and animals however do not learn complex tasks in this way. For instance, human infants learn to speak by first going through a “babbling” stage as they learn the “feel” or the range of their articulatory commands. They also listen carefully to the speech around them, initially learning it implicitly without necessarily producing any of it. When infants learn to speak, they utter incomplete malformed replicas of what they hear. They also sense these errors (unsupervised) or are told about them (semi-supervised) and proceed to adapt the articulatory commands to minimize the errors and slowly converge on the desired auditory signal. In other words, learning these complex sensorimotor mappings proceeds simultaneously and often in an unsupervised manner by listening and speaking all at once \cite{mirrorNetpaper, bird_paper}.


Motivated by such learning of complex sensorimotor tasks, an autoencoder architecture, the “Mirror Network” (or MirrorNet), was proposed in \cite{mirrorNetpaper}. The essence of this biologically motivated algorithm is the bidirectional flow of interactions (`forward' and `inverse' mappings) between the auditory and motor responsive regions, coupled to the constraints imposed simultaneously by the actual motor plant to be controlled. Siriwardena et al. \cite{yashish_music} extended this work to demonstrate the efficacy of the MirrorNet architecture in learning audio synthesizer controls to produce a given melody of notes in a completely unsupervised fashion. In our current work, we explore using the same MirrorNet architecture to learn articulatory representations from a given speech input by incorporating a custom developed DNN based articulatory synthesizer.

\vspace*{-6pt}
\subsection{Background and the Expected Goals}
\vspace*{-3pt}
\label{ssec:previous work}

The idea of determining articulatory parameters to control a vocal tract model was explored in \cite{Yoshikawa_mirror, Warlaumon_nn_mirror, chen21m_interspeech} to simulate the vocal-motor development in infants to learn to control the process of phonation. Similar approaches have recently been explored in \cite{sun22b_interspeech, Georges_mirror, beguvs2022articulation} to do acoustic-to-articulatory speech inversion with parametric vocal tract models \cite{vocaltractlab} or DNN based articulatory synthesizers. Majority of these work focus on how well the input speech is re-synthesized and/or fail to show strong quantitative results to verify the similarity between the learned articulatory representations and the ground-truth. Some of the work here are also limited to learning articulatory representations to a given set of words or phonemes \cite{beguvs2022articulation, westerman_mirrorneurons}. 

The goals we try to achieve with the proposed MirrorNet learning algorithm are two fold; We want to re-synthesize unseen, continuous speech utterances by driving a speaker-independent articulatory synthesizer. In doing that, we want to learn meaningful articulatory representations that can be beneficial in understanding how the speech was produced (i.e how the constriction degree and location of articulators evolved).

\vspace*{-5pt}
\section{The Articulatory Synthesizer}
\vspace*{-4pt}
\label{sec:tv_synthesizer}


We begin by describing the Articulatory Synthesizer that will be later embedded into the MirrorNet. Previous work has elaborated that articulatory parameters or gestures can be used to synthesize continuous, co-articulated and intelligible speech which can replicate realistic models of the vocal tract \cite{Maeda1990, birkholz2006construction}.  Articulatory based speech synthesizers can be mainly divided into two categories: (i) Geometrical approaches which model the geometry of the vocal tract \cite{toutios2011estimating, story2013phrase}, and, (ii) Machine learning based \cite{bocquelet14_interspeech, wu22i_interspeech} which learn the non-linear relationships between articulatory representations and acoustic data. In this work, we focus on developing a DNN based articulatory synthesizer trained and evaluated on a publicly available articulatory dataset.




\vspace*{-6pt}
\subsection{Dataset Description}
\vspace*{-3pt}
\label{sec:dataset}

We use 4 hours of speech data from 46 speakers (21 males, 25 females) of the U. of Wisconsin XRMB database \cite{Westbury1994b}. Using geometric transformations, the XRMB trajectories (X-Y positions of the pellets movement) were converted to TV trajectories as outlined in \cite{Mitra2012}. The transformed database comprises of six TV trajectories: Lip Aperture (LA), Lip Protrusion (LP), Tongue Body Constriction Location (TBCL), Tongue Body Constriction Degree (TBCD), Tongue Tip Constriction Location (TTCL) and, Tongue Tip Constriction Degree (TTCD).

The XRMB dataset was divided into training$_{full}$, development, and testing splits, so that the training$_{full}$ set has utterances from 36 speakers and the development and testing sets have 5 speakers each (3 males,2 females). A subset of the subjects (4 speakers) from the training$_{full}$ split is used to create an initialization split. The resulting training split, training$_{red}$ (after subtracting the data from 4 speakers) along with the initialization split are used in the experiments in sections \ref{ssec:full_vs_light_syn} and \ref{ssec:initialization}. None of the training$_{full}$, development and testing sets have overlapping speakers and hence all the models are trained in a `speaker-independent' fashion. 

\vspace*{-8pt}
\subsection{Proposed synthesizer architecture and model training}
\vspace*{-2pt}
\label{ssec:synth_archi}

We developed a Temporal Convolution Network (TCN) based articulatory speech synthesizer to learn the mapping from six TVs + source features (aperiodicity, periodicity and pitch extracted from \cite{APPdetector}) to the target auditory spectrograms. The model is optimized using the Mean Squared Error (MSE) loss computed between the predicted and the true auditory spectrograms of the input utterance. The resulting auditory spectrograms are then inverted using the algorithm implemented in \cite{Yang_spec_recons} to generate the final acoustic signals.


The TV based synthesizer is implemented in PyTorch with 1-D CNN layers. The complete network is inspired by the multilayered TCN \cite{Lea_2017_Temporal_CNN} and has an identical architecture to the decoder in the MirrorNet (described in section \ref{ssec:model_impl}). To train the synthesizer, learning rates were determined from a grid search by testing all combinations from [1e-2, 1e-3, 1e-4] that resulted in 1e-3 as the best pick. A similar grid search was done to choose the batch size from [16, 32, 64, 128] and 16 gave the best validation MSE. The objective function was optimized using the ADAM optimizer with an `ExponentialLR' learning rate scheduler and a decay of 0.5 (monitoring the validation loss). Sample audio reconstructions can be found in the web page\footnote[1]{https://yashish92.github.io/MirrorNet-for-speech/}


\vspace*{-7pt}
\subsection{Importance of adding source level features}
\vspace*{-3pt}
\label{ssec:6tvs_9tvs_comp}

To investigate the importance of incorporating source level features for articulatory synthesis, we trained two synthesizers with one using 6 TVs and the other using 6 TVs + source features. The two synthesizers when evaluated on the test split have MSEs of 1.6493(0.23) and 2.0607(0.31), respectively. The auditory spectrograms (a), (b) and (c) in figure \ref{fig:tv_synth_results} corresponds to the original speech utterance, the synthesized spectrogram by the proposed articulatory synthesizer trained with 6 TVs + source features and the one synthesized with only 6 TVs respectively. The auditory spectrograms clearly show that the synthesizer trained with source level features generates a better harmonic structure. Hence, for subsequent experiments, source features are used as inputs to the synthesizer.

\vspace*{-7pt}
\subsection{Fully trained vs lightly trained synthesizer}
\vspace*{-3pt}
\label{ssec:full_vs_light_syn}

\begin{figure}[th]
    \centering
    \includegraphics[width=\linewidth, height=48mm]{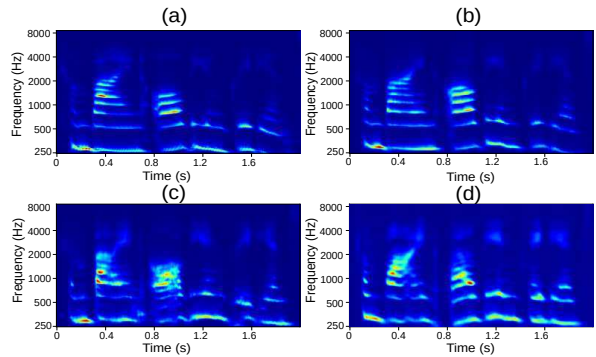}
    \vspace*{-14pt}
    \caption{Auditory spectrogram outputs from the articulatory synthesizers. (a) Input speech utterance, (b) FT synthesizer with source features, (c) FT synthesizer 'without' source features, (d) LT synthesizer with source features}
    \label{fig:tv_synth_results}
    \vspace*{-12pt}
\end{figure}

Since the goal of the MirrorNet is to learn articulatory representations in a completely unsupervised or semi-supervised fashion with minimal exposure to ground-truth articulatory data, it can be expected that the articulatory synthesizer may also have to be trained with a limited amount of ground-truth data based on availability. To test the impact of using a fully trained articulatory synthesizer vs a lightly trained articulatory synthesizer as the vocal tract model in the MirrorNet, two versions of articulatory synthesizers were trained, (i) fully trained (FT): train$_{full}$ split (36 speakers, ~3hours), (ii) lightly trained (LT): Initialization split (4 speakers, ~30min). Here the FT synthesizer is trained with the same configurations in \ref{ssec:synth_archi} whereas the LT synthesizer has the same architecture and configurations except it converged better with a batch size of 64. Both the synthesizers were evaluated on the same original test split after training. Auditory spectrogram (d) in figure \ref{fig:tv_synth_results} shows the output from the LT synthesizer whereas (b) shows the auditory spectrogram from the FT synthesizer for the same input articulatory parameters. The two synthesizers when evaluated on the test split have mean MSEs of 2.1975(0.04) and 1.6493(0.02), respectively. 


\vspace*{-2pt}

\vspace*{-5pt}
\section{MirrorNet Implementation}
\label{sec:ModelArchitecture}
\vspace*{-2pt}

\subsection{Model Architecture and Learning Phase}
\vspace*{-2pt}
\label{ssec:model_archi}

The MirrorNet was initially proposed as a model for learning to control the vocal tract and is based on an autoencoder architecture \cite{mirrorNetpaper}. The goal of the model is to learn two neural projections, an inverse mapping ($\phi$) from auditory representation to motor parameters (Encoder), and a forward mapping ($f$) from the motor parameters to the auditory representation (Decoder). The current MirrorNet implementation is an extension of the work in \cite{yashish_music} where the motor plant is replaced with a DNN based articulatory synthesizer ($g$). The goal of the proposed network is to estimate articulatory representations as the latent space of the autoencoder. We use auditory spectrograms \cite{Wang_auditory_spec} as the input and output representations of the utterances. The auditory spectrograms have a logarithmic frequency scale to provide a unified multi-resolution representation of the spectral and temporal features likely critical in the perception of sound \cite{Wang_auditory_spec}. 


For a given input auditory spectrogram $x_{i} \epsilon R^{C\times L}$ where $C=128$ and $L=250$, the encoder generates a latent space, $\hat{l}$ such that $\hat{l} = \phi(x_{i})$. The decoder then generates an auditory spectrogram $x_{d}$ from the estimated $\hat{l}$ such that $x_{d}=f(\hat{l})$. The synthesizer which takes in the same $\hat{l}$ outputs an auditory spectrogram $x_{s}$ such that $x_{s}=g(\hat{l})$. The MirrorNet model is optimized simultaneously with two loss functions during the `learning phase', namely the `encoder loss'($e_{c}$) and the `decoder loss'($e_{d}$). Here $e_{c} = MSE(x_{d}, x_{i})$ where $x_{d}, x_{i} \epsilon R^{C\times L}$ and $e_{d} = MSE(x_{s}, x_{d})$ where $x_{s}, x_{d} \epsilon R^{C\times L}$. The encoder loss is the typical autoencoder loss whereas the `decoder loss' constrains the latent space to converge to the expected articulatory representation while simultaneously reducing $e_{c}$.  

The role of the `forward' path, $f$ in the MirrorNet is to provide a "neural" model of the synthesizer that runs along the parallel path. Being a neural pathway, it can also back-propagate the errors from the output layers to the control (hidden) layers and subsequently to the earlier Encoder stage, and hence learn the `inverse' mapping, $\phi$ to estimate the latent representation. In general, directly computing the inverse mapping of a vocal-tract or an audio synthesizer is very difficult if not impossible because of its complexity, nonlinearity, and our incomplete knowledge of its workings. MirrorNet solves this problem by adding the neural forward projection as described above, and helps back-propagates the encoder error $e_{c}$ to learn the inverse.   

\begin{figure}[t]
  \centering
  \includegraphics[width=1.0\linewidth,height=34mm]{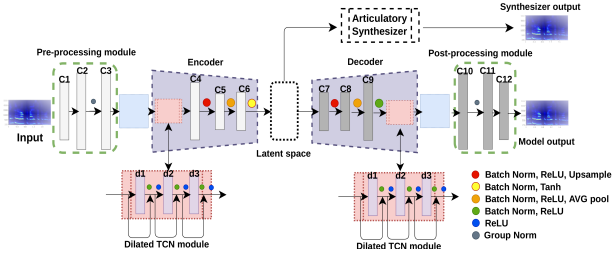}
  \caption{DNN architecture of the MirrorNet model}
    \label{fig:DNN_model}
    \vspace*{-20pt}
\end{figure}

\vspace*{-7pt}
\subsection{Model Implementation and Training}
\label{ssec:model_impl}
\vspace*{-4pt}

The encoder and decoder of the MirrorNet is implemented in PyTorch with 1-D convolutional (CNN) layers. The complete network is inspired by the multilayered Temporal Convolution Network (TCN) \cite{Lea_2017_Temporal_CNN}. Figure \ref{fig:DNN_model} shows the complete DNN model architecture with its sub-modules used for pre/post processing and dilated TCN. The pre/post processing modules are symmetrically matched (C1$\equiv$C12, C2$\equiv$C11, C3$\equiv$C10) and have 128, 256 and 256 filters, respectively, with 1$\times$1 kernels. The d1, d2 and d3 dilated CNN layers have a kernel size of 3 with 1,4 and 16 dilation rates respectively. The CNN layers in the encoder and decoder are also symmetrically matched and the C4, C5 and C6 layers have 256, 128 and 7 filters respectively, with 1$\times$1 kernels. The dimensions of the latent space, $\hat{l}$ are chosen to match with the number of articulatory parameters to be learned and the length of the input speech utterance. For example to learn 9 articulatory parameters sampled at 100Hz for a 2 seconds long speech utterance, we use a latent space of (9$\times$200) dimensions. Upsampling (window size 4) and average pooling (window size 5) are done after C4 and C5 layers, respectively, in the encoder, while upsampling (window size 5) and average pooling (window size 4) is done after C7 and C8 layers, respectively, in the decoder. The final model architecture has around 7.5 million trainable parameters.

Unlike a regular autoencoder, the MirrorNet is trained in two alternating stages in each iteration. The decoder is trained first (to minimize $e_{d}$) for a chosen number of epochs. Then, the encoder is trained (to minimize $e_{c}$) for a given number of epochs and this alternation is continued until both losses converge. The number of iterations of training is decided by monitoring the validation losses computed over the development split. Hyperparameter tuning was also done based on the validation losses at training. The best learning rates for the encoder and decoder, and the batch sizes were determined with a grid search, testing all combinations from [1e-3, 1e-4, 3e-4, 1e-6] for learning rates and [16,32,64,128] for the batch sizes. The best performing models had a learning rate of 1e-6 (for both encoder and decoder) and a batch size of 16. The two objective functions were optimized using the ADAM optimizer with an `ExponentialLR' learning rate scheduler and a decay of 0.5. All the models were trained using NVIDIA Quadro P6000 GPUs and took around 10-11 hours on average for convergence.

\vspace*{-5pt}
\section{MirrorNet with the Articulatory Synthesizer}
\vspace*{-4pt}
\label{sec:tv_synthesizer}

We used the FT articulatory synthesizer in the MirrorNet to learn how to compute from any speech utterance the 6 TVs and source parameters needed for the synthesizer. These articulatory functions are estimated as the latent space of the MirrorNet. To evaluate how well the TVs are estimated by the MirrorNet, we calculate the Pearson Product Moment Correlation (PPMC) score between the estimated and ground truth TVs. During training of the MirrorNet, the already trained articulatory synthesizer weights are frozen and no error is back-propagated through the synthesizer.


\begin{table*}[t]
    \centering
    \LARGE
    \caption{PPMC scores (mean and .std across 6 trials) for articulatory variable prediction. Here 'init' refers to `initialization phase'}
    \vspace{-8pt}
    \label{tab:ppmc_scores_init}
    \resizebox{\textwidth}{!}{
    \begin{tabular}{|l|l|l|l|l|l|l|l|l|l|l|l|}
    \hline
    \textbf{Model}     & \textbf{LA} &\textbf{LP} &\textbf{TBCL} & \textbf{TBCD} &\textbf{TTCL} &\textbf{TTCD} & \textbf{Ap.} &\textbf{Per.} &\textbf{Pitch} &\textbf{AVG. 6TVs} &\textbf{AVG. all}\\ \hline
      MirrorNet(no init)  &0.2033(0.02) &0.4907(0.01) &0.4967(0.03) &0.4760(0.02) &0.4735(0.03)  &0.5114(0.04) &0.5354(0.02)  &0.5143(0.01) &0.5930(0.03) &0.4420(0.02) &0.4771(0.04)
\\ \hline
     MirrorNet(init) &0.7701(0.11) &0.8078(0.03) &0.8132(0.05) &0.8258(0.02) &0.8696(0.06) &0.8783(0.05) &0.8970(0.01) &0.9045(0.03) &0.9125(0.02) &\textbf{0.8286(0.02)} &0.8540(0.03)
 \\\hline
    BiGRNN \cite{yashish_bigrnn}  &0.8801(0.04) &0.6200(0.02) &0.8580(0.03) &0.7382(0.01) &0.6922(0.04) &0.9206(0.01) &- &- &- &0.7848(0.02) &-
\\ \hline
    \end{tabular}}
    \vspace{-15pt}
\end{table*}

\begin{figure}[th]
    \centering
    \includegraphics[width=\linewidth, height=40mm]{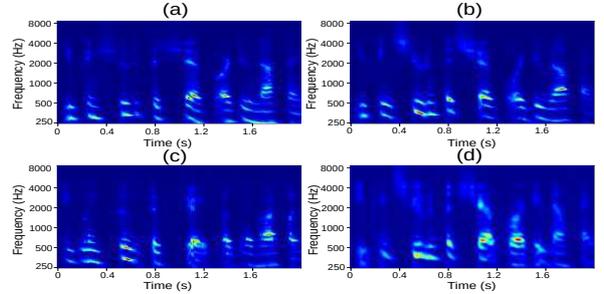}
    \vspace*{-14pt}
    \caption{Auditory spectrogram outputs from the articulatory synthesizers (a) Input speech utterance, (b) Output of FT synthesizer from MirrorNet with init phase, (c) Output of FT synthesizer from MirrorNet without init phase, (d) Output of LT synthesizer from MirrorNet with init phase}
    \vspace*{-13pt}
    \label{fig:mirrornet_specs}
\end{figure}

\begin{figure}[th]
    \centering
    \includegraphics[width=\linewidth, height=70mm]{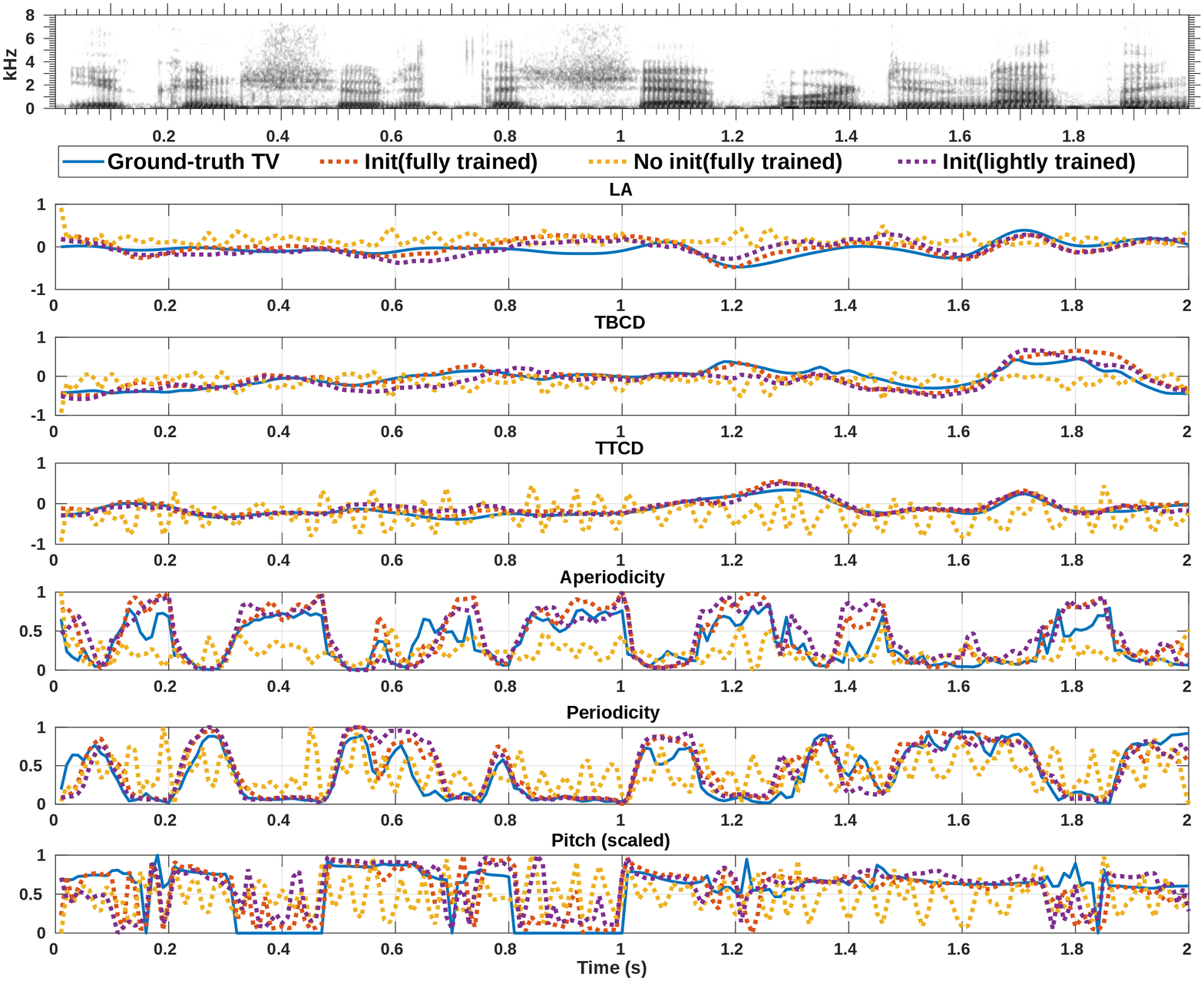}
    \vspace*{-10pt}
    \caption{LA and constriction degree TVs + source features for the utterance ‘You can shoot at the ship or do nothing’ estimated by the MirrorNet. Solid blue Line - ground truth, red dotted line - estimated by the MirrorNet trained with init phase (with FT synthesizer), yellow dotted Line - estimated by the MirrorNet trained without init phase (with FT synthesizer), purple dotted line - estimated by the MirrorNet trained with init phase (with LT synthesizer)}
    \vspace*{-8pt}
    \label{fig:tv_plots_SI}
\end{figure}

\vspace*{-8pt}
\subsection{Learning `meaningful' articulatory representations}
\vspace*{-5pt}
\label{ssec:initialization}

In the previous implementation of the MirrorNet with an audio synthesizer \cite{yashish_music}, the encoder and decoder weights are randomly initialized at the start of training. The latent space converges to a set of parameters (unsupervised) that will best synthesize the input audio (a melody of notes). The audio synthesizer used in \cite{yashish_music} is a parametric model and usually has a unique mapping from the synthesized audio to the corresponding parameters. Therefore, the latent space of the MirrorNet is more constrained and converges fairly easily to the expected parameters. However, the DNN based articulatory synthesizer learns a non-linear mapping from articulatory variables to the auditory spectrograms and, therefore, when used in the MirrorNet could result in non-unique latent space representations. While the quality of the synthesized (final output) speech may well be excellent, it may often converge to non-physiological latent representations, and since the goal of this work is to learn interpretable and meaningful articulatory parameters, we explored here how to \textit{initialize} the encoder and decoder learning to coax the MirrorNet learning to converge eventually to the physiological ranges expected from experimentally measured articulatory representations.

The `initialization phase' of the MirrorNet is done by training the encoder and decoder independently with a small set of ground truth articulatory data. Here we used the initialization split (4 speakers, $\sim$30 min), a subset from the original train split (train$_{full}$) to initialize the network. The encoder loss $e_{c}^{init} = MSE(l, \hat{l})$, where $l, \hat{l} \epsilon R^{N\times k}$ and decoder loss $e_{d}^{init}=MSE(x_{i}, \tilde{x_{d}})$ where $x_{i}, \tilde{x_{d}} \epsilon R^{C\times L}$ is re-defined for the `initialization phase'. Here $e_{c}^{init}$ uses ground truth articulatory data $l \epsilon R^{9\times200}$ unlike in $e_{c}$ in the `learning phase' of the MirrorNet. Similarly $e_{d}^{init}$ in `initialization phase' is different from $e_{d}$ in `learning phase' in two ways, (i) $e_{d}^{init}$ does not use the articulatory synthesizer generated auditory spectrograms, (ii) $\tilde{x_{d}} = f(l)$, and uses the ground truth articulatory representation $l$ unlike $x_{d}$ which only uses $\hat{l}$, the estimated articulatory representation by the encoder in `learning phase'. The initialization phase also uses a larger learning rate for both the encoder and decoder (1e-3), whereas the learning phase uses a comparatively lower learning rate (1e-6). Both the learning rates were determined by a grid search.





Table \ref{tab:ppmc_scores_init} shows the PPMC scores for the articulatory parameter estimation on the test split. The results are from the best performing MirrorNet models trained with and without the initialization phase. We also present results from a best performing BiGRNN model \cite{yashish_bigrnn} trained in a completely supervised fashion and evaluated on the same test split, as a baseline for comparison. Note that the BiGRNN model in \cite{yashish_bigrnn} has only been trained to predict the 6 TVs as targets. 

The Results in Table \ref{tab:ppmc_scores_init} clearly show the importance of the `initialization phase' to estimate meaningful articulatory representations. This finding is further validated from the estimated articulatory trajectories shown in figure \ref{fig:tv_plots_SI}, where the trajectories predicted from the MirrorNet without initialization deviate significantly from the ground truth. However, these representations when fed to the DNN based articulatory synthesizer are producing auditory spectrograms (plot \ref{fig:mirrornet_specs}(c)) that closely resemble the input speech. As previously explained, this result is mainly due to the non-linear and non-unique nature of the DNN based articulatory synthesizer. The most striking observation from Table \ref{tab:ppmc_scores_init} is that the MirrorNet trained with an `initilaization phase' results in a noticeable improvement in TV estimation compared to the BiGRNN speech inversion system. This may be due to the fact that the auditory spectrograms used in the MirrorNet as input representations contain valuable source level information that are lacking in the 13 MFCCs used in the conventional speech inversion system, coupled with the prediction of source level features as additional targets to 6TVs.

\vspace*{-8pt}
\subsection{Semi-supervised vs pseudo semi-supervised modeling}
\vspace*{-2pt}
\label{ssec:semi_vs_pseudo}

The MirrorNet model discussed so far was either trained in a completely unsupervised fashion (random initialization) or in a semi-supervised fashion (tailored initialization). To simulate an actual scenario of having limited ground truth articulatory data, the LT synthesizer in section \ref{ssec:full_vs_light_syn} was used as the vocal tract model and the resulting network is trained with the `initialization' phase. The same initialization split is used to initialize the MirrorNet, and has been used to train the LT version of the synthesizer. Table \ref{tab:ppmc_full_vs_light} shows the PPMC scores for the articulatory variable prediction from the MirrorNet using the FT synthesizer (pseudo semi-supervised) vs the LT synthesizer (semi-supervised). It is remarkable to see that the MirrorNet with the LT synthesizer is doing comparably well in terms of predicting the articulatory variables with respect to the model using the FT synthesizer. However, auditory spectrograms (b) and (d) in figure \ref{fig:mirrornet_specs} shows that the FT synthesizer is producing better quality speech than the LT synthesizer.

\begin{table}
\centering
\normalsize
\caption{PPMC scores (mean and .std across 6 trails) for MirrorNet using the fully trained vs lightly trained synthesizers}
\vspace*{-8pt}
\label{tab:ppmc_full_vs_light}
\resizebox{\columnwidth}{!}{
\begin{tabular}{|l|l|l|l|l|}
\hline
\textbf{Model}     & \textbf{Average (6 TVs)}  & \textbf{Average (all)}\\ \hline
pseudo semi-supervised       &0.8286(0.02)            &0.8540(0.03)  \\ \hline
semi-supervised      &0.8031(0.01)            &0.8219(0.02)  \\ \hline

\end{tabular}}
\vspace*{-20pt}
\end{table}

\vspace*{-8pt}
\section{Discussion and Conclusion}
\label{sec:discus}
\vspace*{-4pt}

In this work, we explored a constrained autoenecoder architecture inspired by sensorimotor interactions, to learn meaningful articulatory representations. A DNN based articulatory synthesizer was custom developed and trained in a speaker-independent fashion to be used as the vocal tract model in the MirrorNet. When incorporated with this articulatory synthesizer, the MirrorNet estimates the 6TVs and source features as the latent space in a semi-supervised fashion. Including an `initialization phase' followed by conventional `learning' procedures resulted in the best predictions of articulatory variables. `Initialization'  presumably constrained the MirrorNet's encoder and decoder coefficients to converge to the range of values expected from experimentally measured articulatory representations. This suggests that the semi-supervised approach is sufficient to ensure that the computed latent representations are physiologically meaningful. A lightly trained version of the synthesizer was also simulated with the MirrorNet to explore the effects of limited availability of ground-truth data for estimating the articulatory representations. Results demonstrate that the MirrorNet can estimate articulatory representations with considerably better accuracy than previous approaches.  Overall, this highlights the effectiveness and power of the MirrorNet's learning algorithm in enabling to solve the conventional acoustic-to-articulatory speech inversion problem with minimal use of ground-truth articulatory data.

As future work, MirrorNet would be experimented with learning to drive complex, parametric vocal tract models in completely unsupervised fashion. More emphasis will also be made in exploring faster training algorithms and using high-dimensional rich input features.

\vspace{-7pt}
\section{Acknowledgement}
\vspace{-5pt}
This work was supported by the NSF grant IIS1764010

\vspace{-5pt}
\bibliographystyle{IEEEtran}
\bibliography{mybib}

\end{document}